\documentclass[aps,prd,twocolumn,superscriptaddress,showpacs,amsmath,amssymb,amsthm,nofootinbib]{revtex4-2}
\usepackage{amsmath,amssymb}
\usepackage{enumitem}
\usepackage[usenames, dvipsnames]{color}
\usepackage{graphicx}
\usepackage{subcaption}
\usepackage[utf8]{inputenc}
\usepackage{url}
 \usepackage{xcolor}

\usepackage{hyperref}

\usepackage{graphicx}% Include figure files
\usepackage{dcolumn}% Align table columns on decimal point
\usepackage{bm}% bold math
\newcommand{\be}{\begin{equation}}             %:skip:
\newcommand{\ee}{\end{equation}}               %:skip:
\newcommand{\ba}{\begin{eqnarray}}
\newcommand{\ea}{\end{eqnarray}}

\begin{document}

\title{Extremal Kerr--Schild Form}

\author{Mokhtar Hassaine}
\email{hassaine@inst-mat.utalca.cl}
\affiliation{Instituto de Matem\'atica, Universidad de Talca, Casilla 747, Talca, Chile}

\author{David Kubiz\v n\'ak}
\email{david.kubiznak@matfyz.cuni.cz}

\affiliation{Institute of Theoretical Physics, Faculty of Mathematics and Physics,
Charles University, V Holešovičkách 2, 180 00 Prague 8, Czech Republic}

\author{Aravindhan Srinivasan}
\email{srinivasan@math.cas.cz}
\affiliation{Institute of Theoretical Physics, Faculty of Mathematics and Physics,
Charles University, V Holešovičkách 2, 180 00 Prague 8, Czech Republic}

\affiliation{
Institute of Mathematics, Czech Academy of Sciences,
Žitná 25, 115 67 Prague 1, Czech Republic}

\date{November 26, 2024}
%\date{\today}          

\begin{abstract}
We propose a novel ansatz, 
%Kerr--Schild form, 
where the full black hole geometry is written as a linear in mass %, $\Delta M=M-M_E$, 
perturbation of the associated extremal black hole base. Contrary to its `standard' version, the corresponding `extremal Kerr--Schild form' is no longer restricted to special algebraic type spacetimes, and is applicable to numerous black hole solutions with matter, such as the charged Kerr-NUT-(A)dS  spacetimes, 
black holes of $D=5$ minimal gauged supergravity, or the charged dilaton-axion rotating solutions. {This 
ansatz is likely to find its applications in black hole perturbation theory,
shed new light on the CFT description of non-extremal black holes, as well as 
be useful for constructing new exact solutions.}
\end{abstract}

\maketitle

\section{Introduction}
\label{sc:intro}

The `standard' Kerr--Schild form \cite{Kerr:1965wfc, Debney:1969zz, Bini:2010hrs}
\be
g=\eta + \Phi\, l\otimes l\,, \label{KS}
\ee
is a truly remarkable ansatz, { %This is 
especially so, because} one of the most astrophysically relevant solutions of the Einstein equations of the current time \cite{EventHorizonTelescope:2019dse},
a rotating black hole described by the Kerr geometry \cite{Kerr:1963ud}, can be written in this form. In the above, $\eta$ is the flat space base metric, $l$ is a null, geodesic and shearfree vector (both in the full geometry $g$ and the base metric $\eta$), and $\Phi$ is a profile  function which `captures the departure' from flat space. Remarkably, this ansatz effectively linearizes the Einstein equations (with the Ricci tensor $R^a{}_b$ linear in $\Phi$). Moreover, as recently shown, the 
Kerr--Schild form also plays an important role in the so called double copy, where, by `taking square root', it gives rise to solutions of the (non-Abelian) gauge theory \cite{Monteiro:2014cda}.

However, the standard Kerr--Schild ansatz in the form \eqref{KS} is also very restrictive. For example, as shown in \cite{Ortaggio:2008iq} it necessarily requires that the corresponding spacetime is of Weyl type II (or more special) of (higher-dimensional) algebraic classification \cite{Coley:2004jv}. This is rather restrictive, as many known black hole solutions (with matter) have Weyl tensor of general algebraic type. For this reason a number of generalizations of the standard Kerr--Schild ansatz \eqref{KS} have been considered in the literature, ranging from the simplest generalization where the flat base metric is replaced by a maximally symmetric space, e.g. \cite{Gibbons:2004js, Malek:2010mh}, to multi-Kerr--Schild spacetimes \cite{Chen:2007fs}, spacetimes with additional spatial vector \cite{Aliev:2008bh,Ett:2010by, Malek:2014dta}, or `generalized' Kerr--Schild transformation
with unrestricted base space, e.g. \cite{Barrientos:2024uuq, Srinivasan:new}.   While the latter is a rather general construction, that goes beyond the type II restriction, it is very likely that a `random' general base  may not be of physical or mathematical interest. 
%To support this perspective, one can also mention that 
{Moreover, even} 
in the case of the Kerr metric, although it is possible to determine the most general null, geodesic, and shear-free congruences for the background metric $\eta$, only one of these congruences leads to the physical Kerr solution. On the other hand, requiring the seed metric to enjoy the same symmetries as the final state of the black hole, one can  uniquely select both the seed metric and the congruence that lead to the Kerr solution  \cite{Ayon-Beato:2015nvz}.  
It thus 
seems interesting to explore alternative physical criteria for selecting the seed metric.

In this paper we specialize to an extremal black hole base, proposing a novel alternative to the standard Kerr--Schild ansatz, applicable to numerous black hole spacetimes with %possibly 
multiple horizons. Namely,
we consider the following `{\em Extremal Kerr--Schild} (EKS) form':
\be\label{ExtremalKS}
g=g_E+\frac{\Delta M}{\Sigma} l\otimes l\,,
\ee
writing the full spacetime of arbitrary mass $M$ as a linear in $\Delta M=M-M_E$ perturbation of the associated  extremal black hole base metric element $g_E$ with extremal mass $M_E$. Here, as before, $l$ is a null and geodesic vector in both $g$ and $g_E$ and $\Sigma$ is {a corresponding profile function.} %some function.

As we shall see, many black hole spacetimes with matter, that do not admit the standard Kerr--Schild form \eqref{KS}, can be cast in the EKS form \eqref{ExtremalKS}. Moreover, choosing the base to be the extremal black hole geometry has many advantages. In particular, many such black holes represent the `ground BPS states' and allow for a simplified description. For example, their perturbation theory admits the CFT description \cite{Guica:2008mu}. While such a description persists even in the non-extremal case, this is only upon taking (a physically unclear) `near horizon approximation' \cite{Castro:2010fd}. We expect that the EKS form will shed new light on these matters. {It may also 
% as well as 
find its applications in black hole `perturbation theory'  and be useful for constructing new exact (non-extremal) solutions.
}

In the remaining part of this paper, we shall exhibit several examples of spacetimes that can be cast in the EKS  form
{and discuss its implications for the double copy framework.}

%%%%%%%%%%%%%%%%%%%%%%%%%%%%%%%%%%%%%%%
\section{Kerr--Newman-(A)dS solution}
\label{K_N_AdS}
%%%%%%%%%%%%%%%%%%%%%%%%%%%%%%%%%%%%%

{To give our first example, let us %first
 turn to the familiar four-dimensional Kerr--Newman-AdS black hole spacetime,  e.g. \cite{Caldarelli:1999xj}. Because of the presence of the cosmological constant, it does not admit the standard Kerr--Schild form \eqref{KS}. Instead, one has to replace the flat base metric $\eta$ with the corresponding maximally symmetric space $g_{\mbox{\tiny AdS}}$, upon which the (electrically charged) Kerr--Newman-AdS solution can be written as}
\ba
g&=&g_{\mbox{\tiny AdS}}+\Bigl(\frac{2mr}{\rho^2}-\frac{q^2}{\rho^2}\Bigr)l \otimes l\,, \label{KS_KerrNewman}\\
A&=&-\frac{qr}{\rho^2}l\,.
\ea
Here,  using the Boyer--Lindquist-type coordinates, the AdS background metric $g_{\mbox{\tiny AdS}}$ 
reads\footnote{The de Sitter version of the metric is obtained by Wick rotating the cosmological constant, $\ell \to i \ell$.
Interestingly, upon the due Wick rotation of coordinates and parameters, one can cast the Kerr--Newman-AdS geometry as a linear in the cosmological constant perturbation of the associated Kerr--Newman spacetime \cite{Frolov:2017whj}, see also the Appendix.
}
\begin{align} \label{metricKN}
g_{\mbox{\tiny AdS}}=&-\frac{f_0}{\rho^2}\bigl({d}t-a\sin^2\!\theta \frac{{d}\varphi}{\Xi}\bigr)^2+\frac{\rho^2}{f_0}{d}r^2 \nonumber\\
&+\frac{S\sin^2\!\theta}{\rho^2}\Bigl[a{d}t-(r^2+a^2)\frac{{d}\varphi}{\Xi}\Bigr]^2+\frac{\rho^2}{S} {d}\theta^2\,,
\end{align}
with
\ba
f_0&=&(r^2+a^2)\Bigl(1+\frac{r^2}{\ell^2}\Bigr)\,,\quad S=1-\frac{a^2}{\ell^2}\cos^2\!\theta\,,\nonumber\\
\rho^2&=&r^2+a^2\cos^2\theta\,,
\quad \Xi=1-\frac{a^2}{\ell^2}\,,
\ea
and the null and geodesic (shear-free) vector $l$, which is also a double 
Weyl Aligned Null Direction (WAND) \cite{Coley:2004jv}, takes the following form:
\be\label{lKerr}
l=dt-a\sin^2\!\theta \frac{d\varphi}{\Xi} +\frac{\rho^2}{f_0}dr\,.
\ee
For any such black hole, characterized by mass $M=\frac{m}{\Xi^2}$, charge $Q=\frac{q}{\Xi}$, angular momentum $J=Ma$, and the cosmological constant $\Lambda=-3/\ell^2$, there exists an associated extremal Kerr--Newman-(A)dS black hole geometry $g_E$, characterized by the extremal mass parameter $m_E$, determined from
imposing a double root of the following function 
\be
f=(r^2+a^2)\Bigl(1+\frac{r^2}{\ell^2}\Bigr)-2m r+q^2\,,
\ee
that is, $f(r=r_E,m=m_E)=0=\partial_rf(r=r_E,m=m_E)$. Hence, the extremal black hole is
 given by \eqref{KS_KerrNewman}, upon replacing $m$ with $m_E$, so that the original black hole geometry can be written as:
\begin{eqnarray}
g&=&g_{\mbox{\tiny AdS}} +\Bigl(\frac{2m_Er}{\rho^2}-\frac{q^2}{\rho^2}\Bigr)l \otimes l+\frac{2(m-m_E)r}{\rho^2}l \otimes l\nonumber\\
&=&g_E+\frac{\Delta M}{\Sigma}l \otimes l\,,\quad \Sigma=\frac{\rho^2}{2 \Xi^2 r}\,,
\end{eqnarray}
{where we used the fact that $\Delta M=\Delta m/\Xi^2$.}
Note that the derived EKS form looks `more natural' than the original Kerr--Schild \eqref{KS_KerrNewman}: the quadratic appearance of charge was absorbed in the extremal background, and
the full Kerr--Newman-(A)dS geometry of arbitrary mass was expressed as a linear in
$\Delta M=M-M_E$, perturbation of the corresponding extremal Kerr-Newman-(A)dS
black hole.\footnote{Note that one can write the extremal geometry $g_E$ in a form similar to $g_{\mbox{\tiny AdS}}$ in \eqref{metricKN}, upon replacing $f_0$ with that of an extremal Kerr-Newman-(A)dS black hole, $f_E\equiv f(m=mE)$. This can be achieved by the following coordinate transformation:
\be\label{transformationBack}
dt\to dt+Zdr\,,\   d\varphi\to d\varphi+\frac{a\Xi Z dr}{r^2+a^2}\,,\ Z=
\Bigl(\frac{1}{f_E}-\frac{1}{f_0}\Bigr)(r^2+a^2)\,,
\ee
upon which the vector $l$ maintains its form \eqref{lKerr}, with $f_0$ replaced by $f_E$.
}

Before we proceed to other examples, let us briefly comment on the geometric interpretation of the null vector $l$. It is well known that the Kerr-Newman-(A)dS family of spacetimes admits a hidden symmetry, encoded in the so called {\em principal tensor}, e.g. \cite{Frolov:2017kze}. This is a non-degenerate closed conformal Killing--Yano 2-form $h$,  obeying the following equation:
\be\label{PT}
\nabla_{c} h_{ab}=g_{ca}\xi_b-g_{cb}\xi_a\,,\quad \xi^a=\frac{1}{3}\nabla_c h^{ca}\,.
\ee
Such a tensor shapes geometric properties of this family. For example, it guarantees separability of the Hamilton--Jacobi, Klein--Gordon, Dirac, and vector field equations in these backgrounds, restricts the algebraic type of the spacetime to type D, and many more, see \cite{Frolov:2017kze} for details.
Moreover, it turns out that the Kerr--Schild vector $l$ {is an eigenvector of $h$, namely:
\be
h^a{}_b\,l^b=\lambda l^a\,,
\ee
for some spacetime dependent eigenvalue $\lambda$.}
In this sense, the hidden symmetry of the principal tensor `encodes' %relates to
the above Kerr--Schild form.

The above construction can be extended to include the magnetic charge and (much more non-trivially) also the NUT parameter, and thus {applies to} 
the general charged Kerr-NUT-AdS spacetimes in four dimensions \cite{carter1968new}.  While the standard approach when both mass and NUT charges are involved is to consider
a multi-Kerr--Schild form ({for which one has to Wick rotate to unphysical spacetime signature}), e.g. \cite{Chen:2007fs}, in our EKS approach, one can easily absorb the NUT parameter in the extremal black hole background, and again recover the form \eqref{ExtremalKS}. We refer to the Appendix for more details on this important case.

%%%%%%%%%%%%%%%%%%%%%%%%%%%%%%%%%%%%%%%
\section{Minimal SUGRA solution in $D=5$}
%%%%%%%%%%%%%%%%%%%%%%%%%%%%%%%%%%%%%%%%%%%
Let us next turn to the Chong--Cveti{\v c}--L{\" u}--Pope black hole solution of $D=5$ minimal gauged supergravity~\cite{Chong:2005hr} whose (bosonic) action is given by the following Lagrangian
\be
{\cal L}=*(R+\Lambda)-\frac{1}{2}F\wedge *F+\frac{1}{3\sqrt{3}}(F\wedge F\wedge A)\,. \label{SUGRA_action}
\ee
The Chong--Cveti{\v c}--L{\" u}--Pope black hole is of general algebraic type and hence cannot admit the standard Kerr--Schild form. However, as shown in \cite{Aliev:2008bh}, it can be written in the following {(extended Kerr--Schild) form}:
\ba
g&=&g_{\mbox{\tiny AdS}}+\Bigl(\frac{2m}{\rho^2}-\frac{q^2}{\rho^4}\Bigr)l \otimes l +\frac{q}{\rho^2} {l}\otimes_s {\mathsf{m}}\,,\\
A&=& \frac{q}{2\rho^2} l\,,
\ea
where
\ba
g_{\mbox{\tiny AdS}}&=&-\Bigl(1+\frac{r^2}{\ell^2}\Bigr)\frac{f_\theta}{\Xi_a\Xi_b}dt^2-2dr l+\frac{\rho^2}{f_\theta} d\theta^2\nonumber\\
&\quad &+\frac{(r^2+a^2)\sin^2\!\theta}{\Xi_a} d\varphi^2+\frac{(r^2+b^2)\cos^2\!\theta}{\Xi_b}d\psi^2\,,\nonumber\\
f_\theta&=&1-\frac{a^2}{\ell^2}\cos^2\!\theta-\frac{b^2}{\ell^2}\sin^2\!\theta\,, \quad \Xi_a=1-\frac{a^2}{\ell^2}\,,\nonumber\\
&\rho^2=&r^2+a^2\cos^2\!\theta+b^2\sin^2\!\theta\,, \quad \Xi_b=1-\frac{b^2}{\ell^2}\,.
\ea
Apart from the null geodesic vector $l$, which turns out to be a single WAND \cite{Malek:2014dta}, the above ansatz also employs the spatial vector $\mathsf{m}$ orthogonal to $l$:
\ba
l&=&\frac{f_\theta}{\Xi_a\Xi_b}dt-\frac{a\sin^2\!\theta}{\Xi_a}d\varphi-\frac{b\cos^2\!\theta}{\Xi_b}d\psi\,,\nonumber\\
\mathsf{m}&=&\frac{f_\theta}{\Xi_a\Xi_b}\frac{ab}{\ell^2}dt-\frac{b\sin^2\!\theta}{\Xi_a}d\varphi-\frac{a\cos^2\!\theta}{\Xi_b}d\psi\,,
\ea
The horizons are determined from the metric function
\be
f=(r^2+a^2)(r^2+b^2)\Bigl(1+\frac{r^2}{\ell^2}\Bigr)+2abq+q^2-2mr^2\,,
\ee
where $m, q$ are the mass, charge parameters, $\ell$ is the cosmological radius, and $a,b$ are the two rotation parameters.

Finding the corresponding extremal black hole from $f(r_E)=f'(r_E)=0$ with $m=m_E$, for fixed $\{a,b, q, \ell\}$, one can derive the corresponding $m_E$, and can rewrite the above as:
\ba
g&=& g_{\mbox{\tiny AdS}}+\Bigl(\frac{2m_E}{\rho^2}-\frac{q^2}{\rho^4}\Bigr)l \otimes l +\frac{q}{\rho^2} l\otimes_s m
+\frac{2\Delta m}{\rho^2}l \otimes l\nonumber\\
&=&g_E+\frac{\Delta M}{\Sigma}l \otimes l\,,\ \Sigma=\frac{\pi\rho^2(2\Xi_a+2\Xi_b-\Xi_a\Xi_b)}{8\Xi_a^2\Xi_b^2}\,,
\ea
where in the last formula, we have employed the relation between the physical mass $M$ and the mass parameter $m$, as given in \cite{Chong:2005hr}, recovering thus the EKS form \eqref{ExtremalKS}.

Interestingly enough, also in this case the Kerr--Schild vector $l$ can be related to the hidden symmetry of the spacetime. Namely, while the `vacuum'  principal tensor no longer exists, one can find its `torsion generalization',  obeying an analogue of \eqref{PT} with torsion. Here, the torsion is naturally identified with the Maxwell 3-form, $*F$,  present in the spacetime, see
\cite{Kubiznak:2009qi} % Houri:2010fr} % ,Houri:2017tlk}
for the corresponding construction. Similar to the Kerr--Newman case, the  Kerr--Schild vector {$l$ is an  eigenvector of this `generalized principal tensor'. 
}

%%%%%%%%%%%%%%%%%%%%%%%%%%%%%%%%%%%%%%%%%%%%%%
\section{Einstein-Maxwell-Dilaton-Axion}
%%%%%%%%%%%%%%%%%%%%%%%%%%%%%%%%%%%%%%%%%%%%%%%%%%

{Various extensions of the Standard Model lead to black holes that typically carry extra fields and charges. Focusing on (astrophysically relevant) %In the astrophysical 
$D=4$ dimensions one such example is the case of a solution 
of the Einstein--Maxwell-dilaton-axion theory}, 
whose action reads
\begin{eqnarray}
S=\int d^4\!x \sqrt{-g}\Big[R-2\partial_{\mu}\phi\partial^{\mu}\phi-\frac{1}{2}e^{4\phi}\, \partial_{\mu}\kappa\partial^{\mu}\kappa\nonumber\\+e^{-2\phi}F_{\mu\nu}F^{\mu\nu}
+\kappa F_{\mu\nu}{(*F)}^{\mu\nu}\Big]\,,
\end{eqnarray}
where $\kappa$ and $\phi$ are the  axion and the dilaton, respectively. 
In \cite{Clement:2002mb}, the authors showed that this theory admits a stationary solution, given by 
\begin{eqnarray}
\label{EMDA}
&&g=-\frac{f}{r_0r}dt^2+r_0r\Big[\frac{dr^2}{f}+d\theta^2+\sin^2\!\theta\big(d\varphi-\frac{a}{r_0r}dt\big)^2\Big]\,,\nonumber\\
&& A=\frac{\sqrt{2}}{2}\Big(\frac{\rho^2}{r_0r}dt+a\sin^2\theta d\varphi\Big)\,,\nonumber\\
&& e^{-2\phi}=\frac{r_0r}{\rho^2}\,,\quad \kappa=-\frac{r_0 a \cos\theta}{\rho^2}\,,
\end{eqnarray}
where $\rho^2=r^2+a^2\cos^2\!\theta$ and $f=r^2-2mr+a^2$. This solution represents a rotating black hole with mass $M=\frac{m}{2}$ and angular momentum $J=\frac{a r_0}{2}$, see \cite{Clement:2002mb}.

To construct the EKS form \eqref{ExtremalKS} in the previous cases, we have taken a `shortcut', using the fact
that the spacetimes were known to admit a (generalized) Kerr-Schild form. In the case of \eqref{EMDA}, this is not known, and therefore we proceed %slightly 
differently. {Namely, as obvious 
%it is quite evident 
from the expression for $f$, the extremal configuration corresponds to $m=m_E=a$, that is, $f=f_E=(r-a)^2$. Let us denote the corresponding extremal metric $g_E$; it is given  by \eqref{EMDA} with $f$ replaced by $f_E$. At the same time,} it is quite easy to see that the following vector: 
\be 
l=dt+\frac{r_0 r}{f_E} dr
\ee
is null, geodesic, and shear-free with respect to the full and extremal metrics and forms a single WAND. It thus appears to be an ideal candidate for our EKS ansatz \eqref{ExtremalKS}, and in fact, one can show that the stationary non-extremal metric can be recast as
\be
g=g_E+\frac{\Delta M}{\Sigma} \l \otimes l\,,\quad \Sigma=\frac{r_0}{4}\,,
\ee
{with the original form of the metric \eqref{EMDA} recovered upon the following coordinate transformation (similar to \eqref{transformationBack}):
\be
dt\to dt-\frac{2r_0(\Delta m) r^2}{f_Ef}dr\,,\quad d\varphi\to d\varphi -\frac{2a(\Delta m) r}{f_Ef}dr\,,  
\ee
with $\Delta m=m-a=2\Delta M$.
}  

{Another example of this type, which does not allow for the standard Kerr--Schild form \eqref{KS} but admits the EKS form is the Kerr--Sen  geometry~\cite{Sen:1992ua} that arises from the low energy limit of heterotic string theory. While the full details will be discussed elsewhere, let us remark here that even in this case %both these cases 
the corresponding Kerr--Schild vector $l$ can be related to the (generalized) hidden symmetry of the spacetime.  
}

%%%%%%%%%%%%%%%%%%%%%%%%%%%%%%%%%%%%%%%%%%%%%%%%
\section{Comments on Kerr--Schild double copy}

{The embedding of the above solutions in the EKS form has several implications for the double copy framework, which we shall now discuss. In particular, } 
the cases of Kerr, Kerr--Newman, or, more generally, the charged Kerr-NUT-(A)dS solutions discussed in the Appendix admit a test Maxwell field, {given by the following  vector potential:\footnote{Test fields are weak fields whose gravitational backreaction on the spacetime can be ignored. We shall treat such fields as independent from any possible backreacting fields already present in the spacetime.}
    \begin{align}
 {\cal A}_{\mbox{\tiny test} }= \frac{r Q_{\mbox{\tiny test} }}{\rho^2} l\,, \label{test_soln}
    \end{align}
where} $l$ is the corresponding null geodesic Kerr--Schild vector, and $\rho^2$ is as defined in respective cases. Thanks to the geodesic nature of ${l}$, all solutions to the Maxwell equations obtained from a vector potential ${\cal A} \propto l$ decouple from the charge and mass appearing in these metrics \cite{MyersPerry:1986,Ortaggio:2023rzp, Srinivasan:new}. Therefore, the field \eqref{test_soln} is not only a test solution in the full spacetime but can also be considered as one in the extremal background of the EKS  representation. Choosing $Q_{\mbox{\tiny test}}=\Delta M=M-M_E$, we see that ${\cal A}_{\mbox{\tiny test} }= \frac{\Delta M}{\Sigma}l$, thus {gives}  %forming 
examples of Kerr--Schild double copy in curved (extremal) backgrounds \cite{Monteiro:2014cda,Bahjat_double_copy_curved_bg_1,Carrillo-Gonzalez:2017iyj,Prabhu_double_copy_curved}.

 Although the notion of the Kerr--Schild double copy for the examples mentioned in this section is already known in the literature, the EKS form offers some new perspectives. First of all, it is clear that the EKS form of these spacetimes can be regarded as a linear perturbation around an extremal background that is  constructed via the double copy of the 
test Maxwell solution \eqref{test_soln}. Moreover, in contrast to the previously known Kerr--Schild double copy for Taub-NUT \cite{KS_double_copy_NUT}, which utilizes its double Kerr--Schild form \cite{PlebanskiDemianski:1976} (cf. equation \ref{double_KS}), the EKS  form presents a different double copy prescription by employing the single Kerr--Schild splitting of the metric with a curved (extremal) background, as given by \eqref{final} and \eqref{extremal_bg_Kerr_NUT} {in the Appendix}. Finally, for the charged solutions namely Kerr--Newman and charged Kerr-NUT-(A)dS, the EKS form allows for a notion of Kerr--Schild double copy in which the background already includes a backreacting matter field in addition to the test field.

%%%%%%%%%%%%%%%%%%%%%%%
\section{Conclusions}
%%%%%%%%%%%%%%%%%%%%%%%

In this paper, we have proposed the `extremal Kerr--Schild form' \eqref{ExtremalKS}, where the flat base metric of the standard Kerr--Schild ansatz is replaced with an element of the corresponding extremal black hole, and the full metric is written as a linear in $\Delta M$ perturbation of the associated extremal black hole cousin, if it exists.

{
Moreover, we have provided} three concrete examples of black holes with matter, namely the black hole of minimal gauged supergravity, the Einstein-Maxwell-dilaton-axion stationary solution and the 4-dimensional charged Kerr-NUT-AdS spacetime, that do not admit the standard Kerr--Schild form, but we were able to cast them in its extremal version. The first two are of general algebraic type and then go beyond applicability of the standard Kerr--Schild form, which can only apply to spacetimes of Weyl type-II or more special.  The third result can easily be generalized to general Kerr--NUT-AdS spacetimes in all dimensions \cite{Chen:2006xh, Srinivasan:new}. 
%\cite{Chen:2007fs}. 
{In all these cases the corresponding Kerr--Schild vector can be linked to some kind of geometrical hidden symmetry present in these spacetimes and also defines a WAND. } We expect that many more black hole geometries can be cast in the EKS form in this way.

{
On the other hand, the requirement on the existence of the extremal base geometry a priori excludes}  `non-black hole' spacetimes such as plane wave solutions, but also some of the black holes that do not admit their extremal versions, such as the (higher-dimensional) Schwarzschild solution or the singly spinning Myers--Perry black holes in $d\geq 6$ dimensions. While all these spacetimes admit the standard Kerr--Schild form, it is impossible to cast them in the EKS version. In such cases, however, one might be able to `replace' the extremal base with some other convenient associated `reference background', constructing thus the corresponding `reference Kerr--Schild form'. In this context, it seems very natural, for example, to identify the base space with the ground state of a given theory. An illustrative example can be provided by the three-dimensional  AdS soliton \cite{Horowitz:1998ha}
\ba \label{sol3d}
 g_s&=&-f_sdt^2+\frac{dr^2}{f_s}+r^2d\varphi^2\,,\nonumber\\
 f_s&=&r^2-M_s\,,\quad M_s=-1\,,
\ea
 which represents the ground state of the Einstein equations with a negative cosmological constant. The soliton is known to play a particularly significant role in the AdS$_3$/CFT$_2$ correspondence, notably through its direct application in the Cardy formula to calculate the asymptotic density of states \cite{Correa:2011dt}, and reproducing the Gibbons-Hawking formula for the BTZ entropy as done by a direct use of the central charge \cite{Strominger:1997eq}. {Moreover, one can easily check that the  BTZ solution \cite{Banados:1992wn} of arbitrary mass $M$ can be obtained from the soliton metric \eqref{sol3d} by means of the reference Kerr--Schild  ansatz:
 } 
\be
 g_{\mbox{{\tiny BTZ}}}= g_s+\Delta M\Bigl(dt-\frac{dr}{f_s}\Bigr)^2\,,
\ee
where $\Delta M=M-M_s$.\footnote{
The standard static BTZ form can be recovered by performing the following change of variable: 
\be
dt\to dt-\frac{(\Delta M) dr}{f_sf},\,\quad f=r^2-M\,.
\ee
}
Another interesting reference background could be provided by recently studied self-dual solutions   
\cite{Crawley:2021auj}.

Our proposal thus provides a widely applicable tool of recasting various geometries in a new intriguing form. We hope that this will shed some light on perturbations of these black holes, and their CFT description. It may also lead to some extension of the `Kerr--Schild double copy program', as well as be useful for constructing new exact solutions. In particular, one may attempt to use the extremal Kerr--Schild ansatz as a 'recipe' for constructing an unknown non-extremal geometry from a given extremal spacetime. %In this perspective, we
We provide a toy example of such a `derivation' of the full Kerr-NUT geometry starting from the extremal one in the Appendix.
Along the same lines, it would be interesting to see if it is possible to `lift' the extremal near-horizon geometries classified in \cite{Kunduri:2013gce} to first full extremal and then, via our construction, to non-extremal spacetimes, see also \cite{Chrusciel:2012jk, Hollands:2012xy} for related results.

Finally, in our investigations we have fully focused on black holes in Einstein gravity, possibly subject to additional matter fields and charges. It remains to be seen, whether it is possible to extend such a construction to black holes in other (higher curvature) gravities, building on so far (only partially successful) attempts, e.g. \cite{Anabalon:2009kq, Ett:2011fy}.

\appendix

%%%%%%%%%%%%%%%%%%%%%%%%%%%%%%%%%%%%%%%%%%%%%%%%%%%%%%%%%%%%%%%%%%%%
\section{Charged Kerr-NUT-(A)dS spacetimes}\label{Kerr-NUT_charged}
%%%%%%%%%%%%%%%%%%%%%%%%%%%%%%%%%%%%%%%%%%%%%%%%%%%%%%%%%%%%%%%%%%%%

In this appendix, we shall discuss general charged Kerr-NUT-(A)dS spacetimes in four dimensions \cite{carter1968new}. It is well known that in the presence of a NUT charge, such spacetimes do not admit the standard Kerr--Schild form \eqref{KS}, and the `best one can do' is to cast them in a double Kerr--Schild form, upon Wick rotating some of the coordinates and parameters (see below). Here, we review these results, as well as show that the general metric can be cast in the extremal Kerr--Schild form \eqref{ExtremalKS} introduced in the main text. Our discussion follows in places the exposure in \cite{Frolov:2017whj}.

%%%%%%%%%%%%%%%%%%%%%%%%%%%%%%%%%%
\subsection{Canonical metric}
%%%%%%%%%%%%%%%%%%%%%%%%%%%%%%%%%%%
We start from the following Carter's canonical form  for general charged Kerr-NUT-(A)dS spacetimes \cite{carter1968new}:
\ba\label{canonicalmetric}
g&=&-\frac{f}{\rho^2}(d\tau+y^2d\psi)^2+\frac{f_y}{\rho^2}(d\tau-r^2 d\psi)^2\nonumber\\
&&\quad +\frac{\rho^2}{f}dr^2+\frac{\rho^2}{f_y}dy^2\,,\nonumber\\
A&=&-\frac{\mathsf{e}r}{\rho^2}(d\tau+y^2 d\psi)-\frac{\mathsf{g}y}{\rho^2}(d\tau-r^2 d\psi)\,,
\ea
with the metric functions given by:
\ba
f&=&-c_0+c_1 r^2-\frac{1}{3}\Lambda r^4-2mr+q^2\,,\nonumber\\
f_y&=& -c_0-c_1 y^2-\frac{1}{3}\Lambda y^4+2ny\,,\nonumber\\
\rho^2&=&r^2+y^2\,,\quad q=\sqrt{\mathsf{e}^2+\mathsf{g}^2}\,.
\ea
The solution is characterized by two integration constants $c_0$ and $c_1$ (one of which can be gauged away by means of a coordinate shift), the mass parameter $m$, the NUT charge parameter $n$, the electric $\mathsf{e}$ and magnetic $\mathsf{g}$ charges, and the cosmological constant $\Lambda$.\footnote{
In order to write the solution in the Boyer--Lindquist-type coordinates, one would have to introduce the rotation parameter $a$, by setting $c_0=-a^2, c_1=1-a^2\Lambda/3$, as well as introduce new coordinates $(t,\varphi, r, \theta)$ defined as $y=a\cos \theta\,,\quad \psi=\varphi/a\,,\quad \tau=t-a\varphi$.}

%%%%%%%%%%%%%%%%%%%%%%%%%%%%%%%%%%%%%%%%%%%%%%%%%%%
\subsection{Extremal Kerr--Schild form}
%%%%%%%%%%%%%%%%%%%%%%%%%%%%%%%%%%%%%%%%%%%%%%%%%%%

In order to write the above solution in the extremal Kerr--Schild form, we introduce a null vector
\be\label{l_appendix}
l=d\tau+y^2d\psi+\frac{\rho^2}{f}dr\,,
\ee
and write
\ba
g&=&-\frac{f}{\rho^2}\Bigl[(d\tau+y^2d\psi)^2-\frac{\rho^4}{f^2}dr^2\Bigr]+\frac{f_y}{\rho^2}(d\tau-r^2 d\psi)^2\nonumber\\
&&\quad +\frac{\rho^2}{f_y}dy^2\,\\
&=&-\frac{f}{\rho^2}l^2+2dr l+\frac{f_y}{\rho^2}(d\tau-r^2d\psi)^2+\frac{\rho^2}{f_y}dy^2\,.\label{dva}
\ea
Changing now coordinates according to
\be%\label{CTCarter}
d\tau\to  d\tau-\frac{r^2}{f}dr\,,\quad d\psi\to d\psi-\frac{dr}{f}\,, \label{Kerr_NUT_coord_transf}
\ee
we find that $(d\tau-r^2 d\psi)$ remains unchanged, so that the expression for the metric $g$, \eqref{dva}, remains formally the same, but with the vector $l$ now given by
\be
l=d\tau+y^2 d\psi\,.
\ee
The metric function $f$ thus now only appears in the first term of \eqref{dva}. Let us further define
\be
f_E\equiv f+2r\Delta m\,,\quad \Delta m\equiv m-m_E\,,
\ee
%\be
%f_0\equiv f+2mr\,,\quad f_E\equiv f_0-2m_Er=f+2r\Delta m\,,
%\ee
where $m_E$ is the extremal black hole parameter (if it exists), derived from $f(r_E)=0=f'(r_E)$ and $m=m_E$ (keeping all other parameters fixed). We may thus rewrite \eqref{dva} as follows:
\ba
g&=&-\frac{f_E-2r\Delta m}{\rho^2}l^2+2dr l+\frac{f_y}{\rho^2}(d\tau-r^2d\psi)^2+\frac{\rho^2}{f_y}dy^2\nonumber\\
&=&g_E+\frac{\Delta m}{\Sigma}l^2\,,\quad \Sigma=\frac{\rho^2}{2r}\,,\label{final}
\ea
recovering formally the extremal Kerr--Schild form \eqref{ExtremalKS}. At the same time, the vector potential reads
\be
 A=-\frac{\mathsf{e}r}{\rho^2}l-\frac{\mathsf{g}y}{\rho^2}(d\tau-r^2 d\psi)\,.
\ee
Note the manifestation of a `single copy' in the absence of magnetic charges, $\mathsf{g}=0$.

Let us have two additional remarks. i) The final form \eqref{final} is written in terms of the mass parameter $m$, not the physical mass itself. Unfortunately, in the presence of NUT charges, the physical mass is currently not completely established. However, if it is proportional to the mass parameter, for example, as suggested in \cite{BallonBordo:2020mcs} for the  special case of vanishing cosmological constant, the above formula can easily be rewritten in terms of the physical mass (see, however, \cite{Liu:2022wku} for a different proposal for the mass of NUTty spacetimes). ii) The extremal base metric
\be
g_E=-\frac{f_E}{\rho^2}l^2+2dr l+\frac{f_y}{\rho^2}(d\tau-r^2d\psi)^2+\frac{\rho^2}{f_y}dy^2\,, \label{extremal_bg_Kerr_NUT}
\ee
while independent of the mass parameter $m$, depends on all other parameters of  the solution, including $n$ and charges  $\mathsf{e}$ and $\mathsf{g}$. This is to be compared with the double Kerr--Schild form discussed below.

%%%%%%%%%%%%%%%%%%%%%%%%%%%%%%%%%%%%%%%%%%%%%%%%%%%%%%%%%%
\subsection{Principal tensor and the Kerr--Schild vector}
%%%%%%%%%%%%%%%%%%%%%%%%%%%%%%%%%%%%%%%%%%%%%%%%%%%%%%%%

The canonical metric \eqref{canonicalmetric} admits a hidden symmetry of the principal Killing--Yano tensor $h$. Explicitly $h$ can be written as, e.g. \cite{Frolov:2017kze}:
\be
h=db\,,\quad 2b=(y^2-r^2)d\tau-r^2y^2 d\psi\,.
\ee

It is easy to verify that this object  is non-degenerate (as a matrix) and obeys the defining equation \eqref{PT}.
Moreover, it can be written as:
\be\label{h222}
 h=-r \mathsf{n} \wedge l+y e_1\wedge e_2\,,
\ee
where $l$ is given by \eqref{l_appendix}, and
\be\label{ne1e2}
\mathsf{n}=-\frac{f}{2\rho^2}l+dr,\quad e_1=\frac{\rho}{f_y}dy, \quad e_2=\frac{f_y}{\rho}(d\tau-r^2  d\psi)\,.
\ee

{It can be seen} that $\mathsf{n}$, by definition, is a null vector normalized as $\mathsf{n} \cdot l = 1$, and likewise, $e_1$ and $e_2$ are orthonormal spatial vectors that are also orthogonal to $l$ and $\mathsf{n}$. Therefore, $l$ and $\mathsf{n}$ form two real eigenvectors, and $w_1=e_1+ie_2$ and $w_2=e_1-ie_2$ form two complex eigenvectors of $h$. {In particular, for $l$ we have }
\be
h^a{}_b \,l^b=\lambda l^b\,,
\ee
with $\lambda=r$.
It is also worth noting that up to an overall scaling factor, $l$ and $\mathsf{n}$ are related by a 'reflection' $\tau \rightarrow -\tau, \psi\rightarrow -\psi$, which is a symmetry of the charged Kerr-NUT-(A)dS spacetime \cite{Ortaggio:2023rzp}.

%%%%%%%%%%%%%%%%%%%%%%%%%%%%%%%%%%%%%%%%%%%%%%%%%%%%%%%%%%%%%%%%%%%%%%%
\subsection{Toy example: full spacetime from its extremal cousin}
%%%%%%%%%%%%%%%%%%%%%%%%%%%%%%%%%%%%%%%%%%%%%%%%%%%%%%%%%%%%%%%%%%%%%%%

Let us now discuss a toy example of how (at least in principle) the extremal Kerr--Schild ansatz \eqref{ExtremalKS} could be used to construct the full Kerr-NUT-(A)dS metric (if it were not known) from its (known) extremal version.

To this purpose, we start from the extremal metric \eqref{canonicalmetric} (setting $\Lambda =0$ and $q^2=0$ for simplicity):
\ba\label{canonicalmetricEEEE}
g_E&=&-\frac{f_E}{\rho^2}(d\tau+y^2d\psi)^2+\frac{f_y}{\rho^2}(d\tau-r^2 d\psi)^2\nonumber\\
&&\quad +\frac{\rho^2}{f_E}dr^2+\frac{\rho^2}{f_y}dy^2\,, \label{extrem_Kerr_NUt}
\ea
where
\be
f_E= -c_0+c_1 r^2-2r{\sqrt{-c_1c_0}}\,. \label{f_e_for_extr_Kerr_NUT}
\ee
%,\quad m_E= \,
The extremal geometry \eqref{extrem_Kerr_NUt}, \eqref{f_e_for_extr_Kerr_NUT} lacks explicit information about the mass parameter. Therefore, knowing only the extremal geometry, it is a priori not obvious how one might reintroduce the mass parameter and thereby construct the full non-extremal geometry. As a possible resolution, we propose to use the extremal Kerr--Schild form \eqref{ExtremalKS} as a 'recipe' to construct the full non-extremal geometry, $g$, as

\be\label{ExtremalKS_app1}
g=g_E+\frac{\Delta m}{\Sigma}{\Tilde l\otimes \Tilde l}\,,
\ee
where $\Tilde l$ is a null vector, $\Sigma=\Sigma(r,y)$ is an unknown metric function, and $\Delta m$ is a constant which will introduce the mass parameter into the metric.

To find the right {KS vector $\Tilde l$}, we take a shortcut here, and assume that it is related to the hidden geometrical symmetry of the extremal Kerr-NUT geometry. This is described by the principal tensor $h$, \eqref{h222}, with $f$ replaced by $f_E$. The corresponding real null eigenvectors are given by $l$, \eqref{l_appendix}, and  its `reflected' counterpart $\mathsf{n}$, \eqref{ne1e2}, both with $f$ replaced by $f_E$.
Since, the geometry is invariant under the `reflection', the two null vectors are equivalent. Therefore, without loss of generality, we choose $l$ to be the Kerr--Schild vector for the extremal Kerr--Schild ansatz \eqref{ExtremalKS_app1}.

Imposing the vacuum Einstein equation,  in particular the equation $R^y{}_y=0$ yields the following linear equation for $\Sigma$:
\be
(r^3+ry^2)\partial_r\Sigma+(y^2-r^2)\Sigma=0\,,
\ee
which implies $\Sigma=\Sigma(\rho^2/r)$. The remaining equations then fix $\Sigma \propto \rho^2/(2r)$. Absorbing the  proportionality constant in $\Delta m$ then yields
\be
\Sigma=\frac{\rho^2}{2r}\,.
\ee
{We thus obtain the full non-extremal geometry, which, in retrospect, can be identified with the Kerr--NUT metric \eqref{canonicalmetric} through a coordinate transformation.}

%%%%%%%%%%%%%%%%%%%%%%%%%%%%%%%%%%%%%%
\subsection{Double Kerr--Schild form}
%%%%%%%%%%%%%%%%%%%%%%%%%%%%%%%%%%%%%%%%%%%%

{For completeness, let us also re-derive the double Kerr--Schild form for these spacetimes. 
To do this, we have to switch} to an unphysical signature, by performing the following Wick rotation:
\be
y=iz\,,\quad f_z=-f_y\,,\quad n=i \hat n\,,\quad \mathsf{g}=i \hat{\mathsf{g}}\,,
\ee
upon which $\rho^2=r^2-z^2$, and the function $f_z$ reads
\be
f_z=c_0-c_1 z^2+\frac{1}{3}\Lambda z^4+2\hat n z\,.
\ee
It is then straightforward to see that after the coordinate transformation \eqref{Kerr_NUT_coord_transf}, the Wick-rotated counterpart of the metric \eqref{dva} takes the form:
\ba
g&=&-\frac{f}{\rho^2}l^2+2dr l-\frac{f_z}{\rho^2}\Bigl[(d\tau-r^2d\psi)^2-\frac{\rho^4}{f_z^2}dz^2\Bigr]\nonumber\\
&=&-\frac{f}{\rho^2}l^2+2dr l-\frac{f_z}{\rho^2}\mathsf{m}^2+2dz \mathsf{m}\,,
\ea
where the two (null) vectors are
\be
l=d\tau-z^2d\psi\,,\quad \mathsf{m}=d\tau-r^2d\psi+\frac{\rho^2}{f_z}dz\,.
\ee
Performing an additional coordinate transformation:
\be\label{transf_App1}
d\tau\to d\tau+\frac{z^2}{f_z}dz\,,\quad d\psi\to d\psi+\frac{dz}{f_z}\,,
\ee
leaves $l$ invariant, while $\mathsf{m}$ now reads
\be
\mathsf{m}=d\tau -r^2 d\psi\,.
\ee
Defining further
\be
f_0\equiv f+2mr-q^2\,,\quad f_{z0}\equiv f_z-2\hat n z\,,
\ee
we may thus write
\ba
g&=&-\frac{f_0-2mr+q^2}{\rho^2}l^2+2dr l-\frac{f_{z0}+2{\hat n} z}{\rho^2}\mathsf{m}^2+2dz \mathsf{m}\,,\nonumber\\
&=&g_{\mbox{\tiny AdS}}+\frac{2mr-q^2}{\rho^2} l^2-\frac{2{\hat n}z}{\rho^2} \mathsf{m}^2\,, \label{double_KS}
\ea
which is the double Kerr--Schild form, where the full solution is written as a linear in `mass and charge square' and linear in `NUT charge' perturbation of the maximally symmetric base space
$g_{\mbox{\tiny AdS}}$, given by
\be
g_{\mbox{\tiny AdS}}=
-\frac{f_0}{\rho^2}l^2+2dr l-\frac{f_{z0}}{\rho^2}\mathsf{m}^2+2dz \mathsf{m}\,.
\ee
The above metric is accompanied by the following vector potential:
\be
 A=-\frac{\mathsf{e}r}{\rho^2}l+\frac{\mathsf{\hat g}z}{\rho^2}\mathsf{m}\,.
\ee
In the limit $q=0$, the `single' copy is still present in the double Kerr--Schild form \cite{KS_double_copy_NUT}; as could be expected, the mass parameter is identified with the electric charge and the NUT parameter with the magnetic one.

Note also that a different splitting
\be
f_K=f +\frac{1}{3}\Lambda r^4\,,\quad
f_{zK}=f_z-\frac{1}{3}\Lambda z^4\,,
\ee
results in the following expression for the metric:
\be
g=g_K+\frac{\Lambda}{3\rho^2}\bigl(r^4 l^2-z^4\mathsf{m}^2\bigr)\,.
\ee
That is, the full charged Kerr-NUT-AdS solution can be written as `linear in $\Lambda$' perturbation of the `charged Kerr-NUT' solution base space $g_K$.

\section*{Acknowledgments}
We would like to thank Marcello Ortaggio for useful discussions.
The work of M.H. is partially supported by FONDECYT grant 1210889.
D.K. is grateful for support from GA{\v C}R
23-07457S grant of the Czech Science Foundation and the Charles University Research Center Grant No. UNCE24/SCI/016. The work of A.S. is supported by the Institute of Mathematics, Czech Academy of Sciences (RVO 67985840). 

%\bibliographystyle{JHEP}
%\bibliography{asa}

%\bibliography{references}

\begin{thebibliography}{10}

\bibitem{Kerr:1965wfc}
R.~P. Kerr and A.~Schild, \emph{{Some algebraically degenerate solutions of
  Einstein\textquoteright{}s gravitational field equations}}, {\emph{Proc.
  Symp. Appl. Math.} {\bfseries 17} (1965) 199}.

\bibitem{Debney:1969zz}
G.~C. Debney, R.~P. Kerr and A.~Schild, \emph{{Solutions of the Einstein and
  Einstein-Maxwell Equations}},
  \href{https://doi.org/10.1063/1.1664769}{\emph{J. Math. Phys.} {\bfseries 10}
  (1969) 1842}.

\bibitem{Bini:2010hrs}
D.~Bini, A.~Geralico and R.~P. Kerr, \emph{{The Kerr-Schild ansatz revised}},
  \href{https://doi.org/10.1142/S0219887810004518}{\emph{Int. J. Geom. Meth.
  Mod. Phys.} {\bfseries 7} (2010) 693}
  [\href{https://arxiv.org/abs/1408.4601}{{\ttfamily 1408.4601}}].

\bibitem{EventHorizonTelescope:2019dse}
{\scshape Event Horizon Telescope} collaboration, K.~Akiyama et~al.,
  \emph{{First M87 Event Horizon Telescope Results. I. The Shadow of the
  Supermassive Black Hole}},
  \href{https://doi.org/10.3847/2041-8213/ab0ec7}{\emph{Astrophys. J. Lett.}
  {\bfseries 875} (2019) L1}
  [\href{https://arxiv.org/abs/1906.11238}{{\ttfamily 1906.11238}}].

\bibitem{Kerr:1963ud}
R.~P. Kerr, \emph{{Gravitational field of a spinning mass as an example of
  algebraically special metrics}},
  \href{https://doi.org/10.1103/PhysRevLett.11.237}{\emph{Phys. Rev. Lett.}
  {\bfseries 11} (1963) 237}.

\bibitem{Monteiro:2014cda}
R.~Monteiro, D.~O'Connell and C.~D. White, \emph{{Black holes and the double
  copy}}, \href{https://doi.org/10.1007/JHEP12(2014)056}{\emph{JHEP} {\bfseries
  12} (2014) 056} [\href{https://arxiv.org/abs/1410.0239}{{\ttfamily
  1410.0239}}].

\bibitem{Ortaggio:2008iq}
M.~Ortaggio, V.~Pravda and A.~Pravdova, \emph{{Higher dimensional Kerr-Schild
  spacetimes}},
  \href{https://doi.org/10.1088/0264-9381/26/2/025008}{\emph{Class. Quant.
  Grav.} {\bfseries 26} (2009) 025008}
  [\href{https://arxiv.org/abs/0808.2165}{{\ttfamily 0808.2165}}].

\bibitem{Coley:2004jv}
A.~Coley, R.~Milson, V.~Pravda and A.~Pravdova, \emph{{Classification of the
  Weyl tensor in higher dimensions}},
  \href{https://doi.org/10.1088/0264-9381/21/7/L01}{\emph{Class. Quant. Grav.}
  {\bfseries 21} (2004) L35}
  [\href{https://arxiv.org/abs/gr-qc/0401008}{{\ttfamily gr-qc/0401008}}].

\bibitem{Gibbons:2004js}
G.~W. Gibbons, H.~Lu, D.~N. Page and C.~N. Pope, \emph{{Rotating black holes in
  higher dimensions with a cosmological constant}},
  \href{https://doi.org/10.1103/PhysRevLett.93.171102}{\emph{Phys. Rev. Lett.}
  {\bfseries 93} (2004) 171102}
  [\href{https://arxiv.org/abs/hep-th/0409155}{{\ttfamily hep-th/0409155}}].

\bibitem{Malek:2010mh}
T.~Malek and V.~Pravda, \emph{{Kerr-Schild spacetimes with (A)dS background}},
  \href{https://doi.org/10.1088/0264-9381/28/12/125011}{\emph{Class. Quant.
  Grav.} {\bfseries 28} (2011) 125011}
  [\href{https://arxiv.org/abs/1009.1727}{{\ttfamily 1009.1727}}].

\bibitem{Chen:2007fs}
W.~Chen and H.~Lu, \emph{{Kerr-Schild structure and harmonic 2-forms on
  (A)dS-Kerr-NUT metrics}},
  \href{https://doi.org/10.1016/j.physletb.2007.09.066}{\emph{Phys. Lett. B}
  {\bfseries 658} (2008) 158}
  [\href{https://arxiv.org/abs/0705.4471}{{\ttfamily 0705.4471}}].

\bibitem{Aliev:2008bh}
A.~N. Aliev and D.~K. Ciftci, \emph{{A Note on Rotating Charged Black Holes in
  Einstein-Maxwell-Chern-Simons Theory}},
  \href{https://doi.org/10.1103/PhysRevD.79.044004}{\emph{Phys. Rev. D}
  {\bfseries 79} (2009) 044004}
  [\href{https://arxiv.org/abs/0811.3948}{{\ttfamily 0811.3948}}].

\bibitem{Ett:2010by}
B.~Ett and D.~Kastor, \emph{{An Extended Kerr-Schild Ansatz}},
  \href{https://doi.org/10.1088/0264-9381/27/18/185024}{\emph{Class. Quant.
  Grav.} {\bfseries 27} (2010) 185024}
  [\href{https://arxiv.org/abs/1002.4378}{{\ttfamily 1002.4378}}].

\bibitem{Malek:2014dta}
T.~M\'alek, \emph{{Extended Kerr-Schild spacetimes: General properties and some
  explicit examples}},
  \href{https://doi.org/10.1088/0264-9381/31/18/185013}{\emph{Class. Quant.
  Grav.} {\bfseries 31} (2014) 185013}
  [\href{https://arxiv.org/abs/1401.1060}{{\ttfamily 1401.1060}}].

\bibitem{Barrientos:2024uuq}
J.~Barrientos, A.~Cisterna, M.~Hassaine and J.~Oliva, \emph{{Revisiting
  Buchdahl transformations: new static and rotating black holes in vacuum,
  double copy, and hairy extensions}},
  \href{https://doi.org/10.1140/epjc/s10052-024-13383-4}{\emph{Eur. Phys. J. C}
  {\bfseries 84} (2024) 1011}
  [\href{https://arxiv.org/abs/2404.12194}{{\ttfamily 2404.12194}}].

\bibitem{Srinivasan:new}
A.~Srinivasan, \emph{{Generalized Kerr-Schild spacetimes in arbitrary
  dimensions}},  \href{https://arxiv.org/abs/in preparation}{{\ttfamily in
  preparation}}.

\bibitem{Ayon-Beato:2015nvz}
E.~Ay\'on-Beato, M.~Hassa\"\i{}ne and D.~Higuita-Borja, \emph{{Role of
  symmetries in the Kerr-Schild derivation of the Kerr black hole}},
  \href{https://doi.org/10.1103/PhysRevD.94.064073}{\emph{Phys. Rev. D}
  {\bfseries 94} (2016) 064073}
  [\href{https://arxiv.org/abs/1512.06870}{{\ttfamily 1512.06870}}].

\bibitem{Guica:2008mu}
M.~Guica, T.~Hartman, W.~Song and A.~Strominger, \emph{{The Kerr/CFT
  Correspondence}},
  \href{https://doi.org/10.1103/PhysRevD.80.124008}{\emph{Phys. Rev. D}
  {\bfseries 80} (2009) 124008}
  [\href{https://arxiv.org/abs/0809.4266}{{\ttfamily 0809.4266}}].

\bibitem{Castro:2010fd}
A.~Castro, A.~Maloney and A.~Strominger, \emph{{Hidden Conformal Symmetry of
  the Kerr Black Hole}},
  \href{https://doi.org/10.1103/PhysRevD.82.024008}{\emph{Phys. Rev. D}
  {\bfseries 82} (2010) 024008}
  [\href{https://arxiv.org/abs/1004.0996}{{\ttfamily 1004.0996}}].

\bibitem{Caldarelli:1999xj}
M.~M. Caldarelli, G.~Cognola and D.~Klemm, \emph{{Thermodynamics of
  Kerr-Newman-AdS black holes and conformal field theories}},
  \href{https://doi.org/10.1088/0264-9381/17/2/310}{\emph{Class. Quant. Grav.}
  {\bfseries 17} (2000) 399}
  [\href{https://arxiv.org/abs/hep-th/9908022}{{\ttfamily hep-th/9908022}}].

\bibitem{Frolov:2017whj}
V.~P. Frolov, P.~Krtous and D.~Kubiznak, \emph{{New metrics admitting the
  principal Killing\textendash{}Yano tensor}},
  \href{https://doi.org/10.1103/PhysRevD.97.104071}{\emph{Phys. Rev. D}
  {\bfseries 97} (2018) 104071}
  [\href{https://arxiv.org/abs/1712.08070}{{\ttfamily 1712.08070}}].

\bibitem{Frolov:2017kze}
V.~P. Frolov, P.~Krtous and D.~Kubiznak, \emph{{Black holes, hidden symmetries,
  and complete integrability}},
  \href{https://doi.org/10.1007/s41114-017-0009-9}{\emph{Living Rev. Rel.}
  {\bfseries 20} (2017) 6} [\href{https://arxiv.org/abs/1705.05482}{{\ttfamily
  1705.05482}}].

\bibitem{carter1968new}
B.~Carter, \emph{A new family of einstein spaces}, {\emph{Physics Letters A}
  {\bfseries 26} (1968) 399}.

\bibitem{Chong:2005hr}
Z.-W. Chong, M.~Cvetic, H.~Lu and C.~N. Pope, \emph{General non-extremal
  rotating black holes in minimal five-dimensional gauged supergravity},
  \href{https://doi.org/10.1103/PhysRevLett.95.161301}{\emph{Phys. Rev. Lett.}
  {\bfseries 95} (2005) 161301}
  [\href{https://arxiv.org/abs/hep-th/0506029}{{\ttfamily hep-th/0506029}}].

\bibitem{Kubiznak:2009qi}
D.~Kubiz\v{n}\'{a}k, H.~K. Kunduri and Y.~Yasui, \emph{Generalized
  {K}illing-{Y}ano equations in $d=5$ gauged supergravity},
  \href{https://doi.org/10.1016/j.physletb.2009.06.037}{\emph{Phys. Lett. B}
  {\bfseries 678} (2009) 240}
  [\href{https://arxiv.org/abs/0905.0722}{{\ttfamily 0905.0722}}].

\bibitem{Clement:2002mb}
G.~Clement, D.~Gal'tsov and C.~Leygnac, \emph{{Linear dilaton black holes}},
  \href{https://doi.org/10.1103/PhysRevD.67.024012}{\emph{Phys. Rev. D}
  {\bfseries 67} (2003) 024012}
  [\href{https://arxiv.org/abs/hep-th/0208225}{{\ttfamily hep-th/0208225}}].

\bibitem{Sen:1992ua}
A.~Sen, \emph{Rotating charged black hole solution in heterotic string theory},
  \href{https://doi.org/10.1103/PhysRevLett.69.1006}{\emph{Phys. Rev. Lett.}
  {\bfseries 69} (1992) 1006}
  [\href{https://arxiv.org/abs/hep-th/9204046}{{\ttfamily hep-th/9204046}}].

\bibitem{MyersPerry:1986}
R.~C. Myers and M.~J. Perry, \emph{Black holes in higher dimensional
  space-times}, \href{https://doi.org/10.1016/0003-4916(86)90186-7}{\emph{Ann.
  Phys. (N.Y.)} {\bfseries 172} (1986) 304}.

\bibitem{Ortaggio:2023rzp}
M.~Ortaggio and A.~Srinivasan, \emph{{Charging Kerr-Schild spacetimes in higher
  dimensions}}, \href{https://doi.org/10.1103/PhysRevD.110.044035}{\emph{Phys.
  Rev. D} {\bfseries 110} (2024) 044035}
  [\href{https://arxiv.org/abs/2309.02900}{{\ttfamily 2309.02900}}].

\bibitem{Bahjat_double_copy_curved_bg_1}
N.~Bahjat-Abbas, A.~Luna and C.~D. White, \emph{{The Kerr-Schild double copy in
  curved spacetime}},
  \href{https://doi.org/10.1007/JHEP12(2017)004}{\emph{JHEP} {\bfseries 12}
  (2017) 004} [\href{https://arxiv.org/abs/1710.01953}{{\ttfamily
  1710.01953}}].

\bibitem{Carrillo-Gonzalez:2017iyj}
M.~Carrillo-Gonz\'alez, R.~Penco and M.~Trodden, \emph{{The classical double
  copy in maximally symmetric spacetimes}},
  \href{https://doi.org/10.1007/JHEP04(2018)028}{\emph{JHEP} {\bfseries 04}
  (2018) 028} [\href{https://arxiv.org/abs/1711.01296}{{\ttfamily
  1711.01296}}].

\bibitem{Prabhu_double_copy_curved}
S.~G. Prabhu, \emph{{The classical double copy in curved spacetimes:
  perturbative Yang-Mills from the bi-adjoint scalar}},
  \href{https://doi.org/10.1007/JHEP05(2024)117}{\emph{JHEP} {\bfseries 05}
  (2024) 117} [\href{https://arxiv.org/abs/2011.06588}{{\ttfamily
  2011.06588}}].

\bibitem{KS_double_copy_NUT}
A.~Luna, R.~Monteiro, D.~O'Connell and C.~D. White, \emph{{The classical double
  copy for Taub\textendash{}NUT spacetime}},
  \href{https://doi.org/10.1016/j.physletb.2015.09.021}{\emph{Phys. Lett. B}
  {\bfseries 750} (2015) 272}
  [\href{https://arxiv.org/abs/1507.01869}{{\ttfamily 1507.01869}}].

\bibitem{PlebanskiDemianski:1976}
J.~F. Pleba\'{n}ski and M.~Demia\'{n}ski, \emph{Rotating charged and uniformly
  accelerated mass in general relativity}, {\emph{Ann. Phys. (N.Y.)} {\bfseries
  98} (1976) 98}.

\bibitem{Chen:2006xh}
W.~Chen, H.~Lu and C.~N. Pope, \emph{{General Kerr-NUT-AdS metrics in all
  dimensions}}, \href{https://doi.org/10.1088/0264-9381/23/17/013}{\emph{Class.
  Quant. Grav.} {\bfseries 23} (2006) 5323}
  [\href{https://arxiv.org/abs/hep-th/0604125}{{\ttfamily hep-th/0604125}}].

\bibitem{Horowitz:1998ha}
G.~T. Horowitz and R.~C. Myers, \emph{{The AdS / CFT correspondence and a new
  positive energy conjecture for general relativity}},
  \href{https://doi.org/10.1103/PhysRevD.59.026005}{\emph{Phys. Rev. D}
  {\bfseries 59} (1998) 026005}
  [\href{https://arxiv.org/abs/hep-th/9808079}{{\ttfamily hep-th/9808079}}].

\bibitem{Correa:2011dt}
F.~Correa, C.~Martinez and R.~Troncoso, \emph{{Hairy Black Hole Entropy and the
  Role of Solitons in Three Dimensions}},
  \href{https://doi.org/10.1007/JHEP02(2012)136}{\emph{JHEP} {\bfseries 02}
  (2012) 136} [\href{https://arxiv.org/abs/1112.6198}{{\ttfamily 1112.6198}}].

\bibitem{Strominger:1997eq}
A.~Strominger, \emph{{Black hole entropy from near horizon microstates}},
  \href{https://doi.org/10.1088/1126-6708/1998/02/009}{\emph{JHEP} {\bfseries
  02} (1998) 009} [\href{https://arxiv.org/abs/hep-th/9712251}{{\ttfamily
  hep-th/9712251}}].

\bibitem{Banados:1992wn}
M.~Banados, C.~Teitelboim and J.~Zanelli, \emph{{The Black hole in
  three-dimensional space-time}},
  \href{https://doi.org/10.1103/PhysRevLett.69.1849}{\emph{Phys. Rev. Lett.}
  {\bfseries 69} (1992) 1849}
  [\href{https://arxiv.org/abs/hep-th/9204099}{{\ttfamily hep-th/9204099}}].

\bibitem{Crawley:2021auj}
E.~Crawley, A.~Guevara, N.~Miller and A.~Strominger, \emph{{Black holes in
  Klein space}}, \href{https://doi.org/10.1007/JHEP10(2022)135}{\emph{JHEP}
  {\bfseries 10} (2022) 135}
  [\href{https://arxiv.org/abs/2112.03954}{{\ttfamily 2112.03954}}].



\bibitem{Kunduri:2013gce}
H.~K. Kunduri and J.~Lucietti, \emph{{Classification of near-horizon geometries
  of extremal black holes}},
  \href{https://doi.org/10.12942/lrr-2013-8}{\emph{Living Rev. Rel.} {\bfseries
  16} (2013) 8} [\href{https://arxiv.org/abs/1306.2517}{{\ttfamily
  1306.2517}}].

\bibitem{Chrusciel:2012jk}
P.~T. Chrusciel, J.~Lopes~Costa and M.~Heusler, \emph{{Stationary Black Holes:
  Uniqueness and Beyond}},
  \href{https://doi.org/10.12942/lrr-2012-7}{\emph{Living Rev. Rel.} {\bfseries
  15} (2012) 7} [\href{https://arxiv.org/abs/1205.6112}{{\ttfamily
  1205.6112}}].

\bibitem{Hollands:2012xy}
S.~Hollands and A.~Ishibashi, \emph{Black hole uniqueness theorems in higher
  dimensional spacetimes},
  \href{https://doi.org/10.1088/0264-9381/29/16/163001}{\emph{Class. Quantum
  Grav.} {\bfseries 29} (2012) 163001}
  [\href{https://arxiv.org/abs/1206.1164}{{\ttfamily 1206.1164}}].

\bibitem{Anabalon:2009kq}
A.~Anabalon, N.~Deruelle, Y.~Morisawa, J.~Oliva, M.~Sasaki, D.~Tempo et~al.,
  \emph{{Kerr-Schild ansatz in Einstein-Gauss-Bonnet gravity: An exact vacuum
  solution in five dimensions}},
  \href{https://doi.org/10.1088/0264-9381/26/6/065002}{\emph{Class. Quant.
  Grav.} {\bfseries 26} (2009) 065002}
  [\href{https://arxiv.org/abs/0812.3194}{{\ttfamily 0812.3194}}].

\bibitem{Ett:2011fy}
B.~Ett and D.~Kastor, \emph{{Kerr-Schild Ansatz in Lovelock Gravity}},
  \href{https://doi.org/10.1007/JHEP04(2011)109}{\emph{JHEP} {\bfseries 04}
  (2011) 109} [\href{https://arxiv.org/abs/1103.3182}{{\ttfamily 1103.3182}}].

\bibitem{BallonBordo:2020mcs}
A.~Ballon~Bordo, F.~Gray and D.~Kubiz\v{n}\'ak, \emph{{Thermodynamics of
  Rotating NUTty Dyons}},
  \href{https://doi.org/10.1007/JHEP05(2020)084}{\emph{JHEP} {\bfseries 05}
  (2020) 084} [\href{https://arxiv.org/abs/2003.02268}{{\ttfamily
  2003.02268}}].

\bibitem{Liu:2022wku}
H.-S. Liu, H.~Lu and L.~Ma, \emph{{Thermodynamics of Taub-NUT and Plebanski
  solutions}}, \href{https://doi.org/10.1007/JHEP10(2022)174}{\emph{JHEP}
  {\bfseries 10} (2022) 174}
  [\href{https://arxiv.org/abs/2208.05494}{{\ttfamily 2208.05494}}].

\end{thebibliography}

\providecommand{\href}[2]{#2}\begingroup\raggedright\endgroup

\end{document}